\documentclass{eurocg12}

\begin{document}
\title{New separation theorems and sub-exponential time  algorithms for 
packing and piercing of fat objects    \
}
\author{Farhad Shahrokhi\\
Department of Computer Science and Engineering,  UNT\\
P.O.Box 13886, Denton, TX 76203-3886, USA\
farhad@cs.unt.edu
}

\date{}
\maketitle
\date{} \maketitle


\begin{abstract}
For $\cal C$ a collection of  $n$  objects in $R^d$, let the packing and 
piercing numbers of $\cal C$, denoted by $Pack({\cal C})$, 
and $Pierce({\cal C})$, respectively, be the largest number of pairwise 
disjoint objects in 
${\cal C}$, and the smallest number of  points in $R^d$ that are common to all
elements of ${\cal C}$, respectively.   
When elements of $\cal C$ are fat objects  of arbitrary sizes, 
  we derive sub-exponential time algorithms for the NP-hard problems 
of computing ${Pack}({\cal C})$ 
and $Pierce({\cal C})$, respectively, that  run in  
$n^{O_d({{ Pack}({\cal C})}^{d-1\over d})}$ 
and $n^{O_d({{ Pierce}({\cal C})}^{d-1\over d})}$ 
time, respectively, and $O(n\log n)$ storage. 
Our main tool which is  interesting in its own way, 
is a new separation theorem.   
The algorithms readily give rise to  polynomial time approximation schemes (PTAS)  
that run in $n^{O({({1\over\epsilon})}^{d-1})}$ time and  $O(n\log n)$ storage.
The results favorably compare with many related best known results. 
Specifically, our separation theorem significantly improves the splitting 
ratio of the previous result of Chan, whereas,  
the sub-exponential time algorithms significantly improve upon
the running times of very recent algorithms  of Fox and Pach for packing of spheres.
\end{abstract}

\section{Introduction and Summary}



An effective tool in the design of divide and
conquer graph algorithms is separation.
The separator theorem of  Lipton and Tarjan \cite{LT},\cite{LT2}
asserts  that any $n$ vertex planar graph can be separated into two
subgraphs with  
a splitting  ratio of $1/3-2/3$, that is, each subgraph having  at most $2n/3$ vertices, 
by removing $O({\sqrt n})$ vertices.
Additional results for graphs and geometric objects
 have been obtained  by Alon et al \cite{A}, Alber and Fiala \cite{AF},  
Miller et al \cite{Mi}, Smith and
Wormald \cite{SW}, Chan \cite{Ch}, Fox and Pach \cite{FP1}
\cite{FP2},\cite{FP3},\cite{FP4}, Fox, Pach and Toth \cite{FPT}, and Shahrokhi \cite{Sh}. 
\subsection{Past Related Results}  
Throughout this  paper ${\cal C}$ denotes a finite collection of subsets of 
$R^d$ of cardinality $n$.
Miller et al \cite{Mi} proved that  
given ${\cal C}$ a set of spheres in $R^d$,  where $d$ is fixed and each point is common to only a constant number of spheres, there is a sphere $S$ in 
$R^d$, so that  at  most 
${d+1\over d+2}n$
of the spheres in  ${\cal C}$ are entirely inside of  $S$, at most 
${d+1\over d+2}n$  are
entirely outside, and at most $O_d(n^{d-1\over d})$  intersect the boundary of  $S$.
Smith and Wormald \cite{SW},  among other  results, derived variations of this fundamental result, 
where  the separator is  a box, and hence, reduced the splitting ratio   
to   $1/3-2/3$
in any dimension  $d$.
Nonetheless, their result required  *disjointness*  assumption  of the spheres,
which  could be weakened to the assumption that
there is a very small overlapping  among the objects.
Since the intersection
graphs of many geometric objects exhibit a large vertex connectivity,
it is impossible to separate them with the removal  of a  small 
number of vertices.  Consequently, researchers,  have focused on separating  
intersection graphs of geometric objects, including spheres, with respect to other measures. 

Alber and Fiala \cite {AF} 
studied the separation properties  of unit discs in  $R^2$
with respect to the area which  coincides with the packing number, in case of unit discs.
For a set ${\cal C}$ of unit discs in $R^2$, they derived a separator 
theorem with a constant splitting ratio,  used it  to compute the 
$Pack({\cal C})$ in 
$n^{O({\sqrt{Pack({\cal C})}})}$ time, and also derived a PTAS
for computing the packing number. 

Chan \cite{Ch} studied the packing and piercing
numbers of fat objects of arbitrary sizes in  $R^d$.
He introduced
the concept of a
measure  on fat objects that coincides with the packing and piercing
numbers, and generalized the separation result of Smith and Wormald
for this measure. In simple words, Chan's beautiful separation  
theorem asserts  that given a set ${\cal C}$ of $n$  fat objects in $R^d$, 
with  a sufficiently large measure $\mu({\cal C})$, there is a cube $R$
so that the measure of  objects inside of $R$ is at most
$(1-\alpha)\mu({\cal C})$,     
the measure of  objects outside  of $R$ is at most
$(1-\alpha)\mu({\cal C})$, and the measure of objects intersecting
$R$ is  
$n^{O_d({{ \mu}({\cal C})}^{d-1\over d})}$, where $\alpha$ is a 
constant whose value is about   
$1\over 2^d+1$.
Chan then used  his separation theorem
to design  Polynomial time Approximation Schemes (PTAS) for packing and piercing problems that run in $n^{O(({1\over \epsilon})^d)}$ time and $O(n)$  space.
Specifically, he improved  the running time of the best previous PTAS for packing problem
that  was due
to Erlebach et al \cite{Erl}.

In  \cite{Sh}, we obtained  a combinatorial measure separation theorem 
for a class of  graphs containing the intersection graphs of nearly fat objects  in $R^d$, 
 and obtained  sub-exponential algorithms and PTAS for packing and piercing number of unit hight rectangles, unit discs, and other related problems.  

Fox and Pach \cite{FP1}, \cite{FP2},
\cite{FP3} have developed a series of new separator theorems
for planar curves and string graphs. These  results highly generalize the
planar separator theorem  and have  striking applications
in combinatorial geometry. 
Very recently, they discovered that 
their methods can be used to solve the maximum independent set problem,
in sub-exponential time, for a variety of geometric  graphs \cite{FP4}. 
Specifically,  they have just reported 
a sub-exponential time algorithm with the running time of 
$2^{O_d\big(n^{d\over d+1}polylog n\big)}$ 
(or $n^{O_d\big(n^{d\over d+1}\big)}$),  
for computing the packing number of spheres.

\subsection{Our Results}
We present
several main results. First, we present an improvement to Chan's separation theorem 
which is obtained by combining some of his ideas with the  original work   
of Smith and Wormald. Specifically, using  a box of aspect 
ratio ratio 2 in $R^d$,  as the separator,
we obtain a  parametric splitting ratio of  $1/3-2/3(1+\epsilon)$, 
for any $0<\epsilon\le {1\over 2}$, for  separating the measure, independent of the dimension. 
 Moreover, we provide an explicit (constant) lower bound for $\mu({\cal C})$, in terms of 
$d$ and $\epsilon$, for obtaining such a suitable separation, 
contrasting the result in \cite{Ch} that was obtained for  *sufficiently large 
values*.
We use this separation theorem to derive 
 sub-exponential time algorithms
for packing and piercing problems, running in 
$n^{O_d({{ Pack}({\cal C})}^{d-1\over d})}$ 
and $n^{O_d({{ Pierce}({\cal C})}^{d-1\over d})}$ 
time, respectively, and $O(n\log n)$ storage. 
Finally, we convert these algorithms to PTAS 
that run in $n^{O({({1\over\epsilon})}^{d-1})}$ time and  $O(n\log n)$ storage.

\section{Preliminaries}
For a closed set $B$ in $R^d$, let $\delta_B$ denote the
boundary of $B$, and note that $\delta_B\subseteq B$. 
Let $\bar B$ denote $R^d-B$, that is, $\bar B$ is 
the set of points  *outside* of $B$.  
A collection ${\cal C}$ of objects in $R^d$ is fat, if  for every $r$ and size $r$
box $R$, we can choose a constant number $c$ of points such that every object
in ${\cal C}$ that intersects $R$ and has size at least $r$ contains one of 
these points \cite{Ch}.  It should be noted that  the class of fat objects  contains spheres, cubes, and boxes with bounded aspect ratios.
\vskip .4cm

Let $\cal C$ be a collection of subsets  of $R^d$, and let
$\mu$ be  a mapping  that assigns non-negative values
to subsets  of $\cal C$. 
Chan \cite{Ch} calls  $\mu$ a  measure, if for any 
${\cal A},{\cal B}\subseteq {\cal C}$ the following hold.

$(i)~$ $\mu({\cal A})\le \mu({\cal B})$, if   ${\cal A}\subseteq {\cal B}$.

$(ii)~$ $\mu({\cal A}\cup {\cal B})\le \mu({\cal A})+\mu({\cal B})$. 

$(iii)~$ $\mu({\cal A}\cup {\cal B})=\mu({\cal A})+\mu({\cal B})$, if
no object in  $\cal A$ intersects an object in $\cal B$.

$(iv)~$ Given any $r>0$ and any size$-r$ box $R$ in $R^d$, if 
every object in $\cal A$ has size at least $R$ and intersects $R$, then 
$\mu({\cal A})\le c$, for a constant $c$. 

$(v)~$  A constant-factor approximation to 
$\mu({\cal A})$ can be computed  in ${|{\cal A}|}^{O(1)}$ time.
Moreover, if $\mu({\cal A})\le b$, then, $\mu({\cal A})$  can be computed
exactly in ${|{\cal A}|}^{O(b)}$ time.

\vskip .3cm

Note that feasible solutions to the packing and piercing problems
gives rise to measures.  
\vskip .4cm

Let ${\cal A}\subseteq{\cal C}$, and let $B$ be  a closed subset of $R^d$. 
Let  ${\cal A}_{B-\delta_B}$  and ${\cal A}_{\bar B}$ 
denote, respectively, the set of all  objects in $\cal A$ that are contained in $B-\delta_B$, or are completely inside of $B$, and the set of all objects 
in ${\cal A}$ 
that are contained in $\bar B$, or are completely outside of $B$, respectively.
Let ${\cal A}_{B}$,  and ${\cal A}_{\delta_B}$, denote the set of all objects 
in ${\cal A}$ that have  their centers in $B$, and the set of all objects in
${\cal A}$ that have a point  
in common with $\delta_B$. 

{\it Aspect ratio} of a box in $R^d$ is the ratio of its longest side to 
its shortest side.  
Chan \cite{Ch} proved the  following separation theorem. 

\begin{theorem}\label{t1}
{\sl
Given a measure $\mu$ satisfying $(i)-(iv)$ and a collection $\cal C$
of $n$ objects in $R^d$ with sufficiently large $\mu({\cal C)}$
there is a  box $R$  with
$\mu({\cal C}_{R-\delta_R}),\mu({\cal C}_{\bar R})\ge\alpha\mu(\cal C)$, and
$\mu({\cal C}_{\delta R})=O_d(n^{{\mu(\cal C)}^{d-1\over d}})$, where
$\alpha$ is some  fixed constant.
Moreover, if $(v)$ is satisfied then, $R$ can be computed  in
polynomial time and linear space.
}
\end{theorem}
\section{The Separation Theorem} 

\begin{theorem}\label{t2}
{\sl
Let $\cal C$ be a set  of $n$   objects   in $R^d$, $d\ge  2$ and let 
$\mu$ be a measure  on $\cal C$ satisfying $(i)-(iv)$, and  
let $0<\epsilon\le{1\over 2}$.
If $\mu({\cal C})\ge {({3.c.d^2.8^d\over \epsilon})}^d$, then, there  is a box $R$  
so that 

$$\mu({\cal C}_{R-\delta R}) \le 
{2\over 3}(1+\epsilon) \mu( {\cal C}), \mu({\cal C}_{\bar R})\le 
{2\over 3}(1+\epsilon)\mu({\cal C}),$$ 
and 
$$\mu({\cal C}_{\delta R})=O_d(n^{{\mu(\cal C)}^{d-1\over d}}).$$ 
}
\end{theorem}

{\bf Proof.} 
Let $B$ be  a minimum volume box  with aspect ratio at most 2
with $\mu({\cal C}_B)\ge ({1+\epsilon\over 3})\mu({\cal C})$, whose side lengths are
$l_1\le l_2\le....\le l_d, l_d\le 2 l_1$.
Let $s$ denote the center of $B$. 
For any  $m$ with $1\le m<2^{1\over d}$,  
let $B_m$ be the box that is the magnified version of $B$ by the magnification
factor $m$. Thus, $B_m$ is a box
of side lengths $l^m_1\le l^m_2\le...\le l^m_d, l^m_d\le 2l^m_1$, that  has center $s$, and  contains $B$. Note that $l^m_i=m l_i<2^{1\over d}l_i$, for $i=1,2,..., d$, and hence the  volume of $B_m$  is
strictly less that 2 times the volume of $B$.
By cutting $B_m$, in the middle of its longest side,  
we can decompose $B$ into two boxes
${B_m^i}, i=1,2$, of aspect ratio at most two,  each having a volume strictly
smaller than volume of $B$. Thus, 
$\mu({\cal C}_{B_m^{i}})<  ({1+\epsilon\over 3}){\mu({\cal C})}$,
$i=1,2$, since $B$ has the minimum volume. Consequently, using $(ii)$, we deduce that 
$\mu({\cal C}_{B_m})<{{2\over 3}(1+\epsilon)}\mu({\cal C})$.
Therefore, 
$\mu({\cal C}_{B_m-\delta B_m})<{{2\over 3}(1+\epsilon)}\mu({\cal C})$.
Next, let $C_m$ be a cube of side length $l^m_d$ having center $s$, 
and note that  area of $\delta_{C_m}$ is 
$2.d{l^m_d}^{d-1}<2^{2d-1\over d}.d.{l_d}^{d-1}<4.d.{l_d}^{d-1}$. 
Now, let $l={l_d\over 8\mu(C)^{1\over d}}$, and note that
$\delta_{C_m}$, and hence $\delta_{B_m}$  can  
be covered with at most $d.{8^{d}}.{\mu(\cal C)}^{d-1\over d}$ 
cubes of size $l$.   
Let ${\cal C}^1$ denote the set of all objects  in $\cal C$ of size at least
$l$, and  let $1\le m<2^{1\over d}$. 
Then, by $(iv)$,  
$\mu({\cal C}^{1}_{\delta B_m})\le c.d.8^{d}.{\mu(C)}^{d-1\over d}$.  
Similarly, let ${\cal C}^2$ denote the set of all objects in ${\cal C}$ of size strictly smaller than $l$. We need the following claim to finish the proof.

{\bf Claim.} Let $1\le m_1<m_2<2^{1\over d}$, with
$m_2-m_1\ge{1\over {\mu(\cal C)}^{1\over d}}$, let
$A_1\in {\cal  C}^2_{\delta{B_{m_1}}}$, and let  
$A_2\in {\cal C}^2_{\delta_{B_{m_2}}}$.
Then, $A_1\cap A_2 =\emptyset$. 

{\bf Justification.} 
 For any $x,y\in R^d$, let   $distance(x,y)$ denote the distance between 
$x$ and $y$.  Note that
$m_2-m_1\ge {1\over {\mu(\cal C)}^{1\over d}}$
implies that for any two points $a\in \delta{B_{m_2}}$ and
$b\in \delta{B_{m_1}}$, we must have 
$distance(a,b)\ge {l_1\over 2{\mu(\cal C)}^{1\over d}}\ge {l_d\over 4{\mu(\cal C)}^{1\over d}}\ge 2.l$.
Assume to the contrary that $A_1\cap A_2\ne\emptyset$, and let $x\in A_1\cap A_2$.
Let $x_1\in A_1\cap \delta{B_{m_1}}$, and let $x_2\in A_2\cap \delta_{B_{m_2}}$.
Note that $distance(x,x_1)<l$ and $distance(x,x_2)<l$, and thus,   
$distance(x_1,x_2)<2.l$ which is a contradiction. $\Box$   
\vskip .4cm

Now use the claim and employ $(ii)$ to conclude that 
$$\sum_{j=0}^{\lfloor {(2^{1\over d}-1){\mu({\cal C})}^{1\over d}}\rfloor}\mu({\cal C}^2_{\delta{B_{1+
{j/{\mu({\cal C})}^{1\over d}}}}})\le \mu({\cal C}).$$

It follows that there is a $j$ so that for  
$m^*=1+{j\over {\mu({\cal C})}^{1\over d}}$, we have 
$\mu({\cal C}^2_{\delta B_{m^*}})\le 
{{\mu(C)}^{d-1\over  d}\over 2^{1\over d}-1}\le 
d^2{\mu(C)}^{d-1\over  d}$, since, ${1\over 2^{1\over d}-1}\ge{1\over d^2}$. 
We conclude that $\mu({\cal C}_{\delta B_{m^*}})\le c.d^28^d.n^{{\mu(\cal C)}^{d-1\over d}}$. Now  let $R=B_{m^*}$, and to finish the proof note that for
$\mu({\cal C})\ge {({3c.d^2.8^d\over \epsilon})}^d$, we have 
$\mu({\cal C}_{\delta R})\le {\epsilon\mu({\cal C}_R)\over 3}$. 
$\Box$.

\vskip .4cm

\section{Sub-Exponential Time Algorithms}
Our next result is the following.

\begin{theorem}\label{t3}
{\sl
Let $\cal C$ be a set of $n$ fat objects  in $R^d$, $d\ge  2$, then 
${Pack}({\cal C})$ and $Pierce({\cal C})$ can be computed in 
$n^{O({ Pack({\cal C})}^{d-1\over d})}$ and  
$n^{O({ Pierce({\cal C})}^{d-1\over d})}$, respectively, and  
$O(n\log n)$ storage. 
}

\end{theorem}
{\bf Proof.} 
We provide the  details for computing packing number; The details relevant
to the computation of piercing number are similar, but slightly different.   
We use   the following recursive algorithm
adopted from \cite{LT2} and tailored to our needs, which we previously utilized
in \cite{Sh}. 
\vskip .4cm

{\it Step 1. 
Determine an approximate solution  $\mu$ to the packing number using $(iv)$.
Choose  a separation parameter $\alpha=(1+\epsilon)$ for Theorem  2, and let $C=C(d)$ 
be the  Lower bound on  $\mu({\cal C})$  in Theorem 2. 
Test using $(v)$ to determine if 
$Pack({\cal C})\le C$. If so, then 
compute a solution in   $O(n^{C})$ time and return it. 
If not, 
proceed to the  recursive  step.}
\vskip .3cm

{\it Step 2 (Recursive Step). 
     Find box $R$ 
by applying Theorem \ref{t2} to $\mu$. For any independent  set of objects
$\cal I$ in ${\cal C}_{\delta B}$ compute  
$Pack({\cal C}_B-N({\cal I}))$,  $Pack({\cal C}_{\bar B}-N({\cal I}))$,  
and   return  
$$\max_{\cal I}\{Pack({\cal C}_B-N({\cal I})+Pack({\cal C}_{\bar B}-N({\cal I}))+ |{\cal I}|\}$$  
where the maximum is taken overall independent sets of objects   $\cal I$ in 
${\cal C}_{\delta B}$.} 

\vskip .5cm

It is easy to verify that the algorithm computes $Pack({\cal C})$ correctly.   
For the running time,  let $T(n,p)$ denote  the execution time
of the algorithm on   an instance  of the problem with $Pack({\cal C})=p$.
Note that,  
$T(n,p)\le n^{O(p^{d-1\over d})} T(n,((1-\alpha)p)$, if  $p\ge b$; Otherwise
$T(n,p)\le O(n^b)$. 
It is not difficult to verify that $T(n,p)=n^{O(p^{d-1\over d})}$ as claimed.
$\Box$

\section{Approximation Algorithms}
The algorithms in the previous section gives rise to  
polynomial time approximation schemes (PTAS) for both of stated problems with  $n^{O({({1\over\epsilon})}^{d-1})}$ time and  $O(n\log n)$ 
storage. 
For the PTAS, one can slightly modify the original divide and conquer
approach  in \cite{LT}.

Specifically, one must first obtain an constant time approximation solution 
to the problem  using $(v)$,  use  it to define the measure $\mu$, and apply 
the separation theorem, or Theorem 2,   to this $\mu$.  
In the  recursive step, the divide and conquer algorithm stops when 
the value of the  measure is 
*small*.  That is, if the value of the measure is  
$O({({1\over \epsilon})}^2)$,  then  an exact solution is computed by the 
application of  our  sub-exponential time  algorithm. 
The claims concerning the  running time, storage, and   
quality of the approximation are easy to verify.


\vskip .4cm
   










\begin{thebibliography}{99}



\bibitem{AG}
Agarwal, P.K., Kreveld, M., Suri, S., Label placement by maximum independent sets in rectangles, 
{Comput. Geometry:Theory and Appl.}
 11(3-4) 209-218, 1998.

\bibitem{AF}
Alber J.,  Fiala J., Geometric Separation and Exact Solutions for the Parameterized Independent Set Problem on Disk Graphs,Journal of Algorithms,
Volume 52 ,  Issue 2, 134 - 151, 2004

\bibitem{A} N. Alon, P. Seymour, and R. Thomas, A separator theorem for graphs
   with an excluded minor and its applications, Proceedings 22nd ACM Symposium on the Theory of Computing, STOC90, 1990, 293-299.





\bibitem{Ch}
Chan T.,
Polynomial-time approximation schemes for packing and piercing fat objects
, Journal of Algorithms,  46(2), 178 - 189,  2003. 

~ 

\bibitem{Erl}
Erlebach T.,  Jansen K. and Seidel E.,
 {
Polynomial-time approximation schemes for geometric graphs}.
{ Proc.  12th  ACM-SIAM Symposium on Discrete Algorithms}
(SODA'01), 671-679,  2001.

\bibitem{FP1}
Fox J., Pach J., 
A separator theorem for string graphs and its applications, 
Combinatorics, Probability and Computing, 2009.

\bibitem{FP2}
Fox J., Pach J., 
Separator theorems and Turán-type results for planar intersection graphs,  Advances in Mathematics 219, 1070-1080, 2008.

\bibitem{FP1}
Fox J.,   Pach J.,
A separator theorem for string graphs and its applications,
 Combinatorics, Probability and Computing 19(2010), 371-390.

\bibitem{FP2}
Fox J., Pach J.,
Separator theorems and Turán-type results for planar intersection graphs,  Advances in Mathematics 219, 1070-1080, 2008.

\bibitem {FP3}Fox J., Pach J., Coloring kk-free intersection graphs of geometric objects in the plane,
Proceedings of the twenty-fourth annual symposium on Computational geometry,
346-354, 2008.


\bibitem{FPT} Fox J, Pach J,  T\'oth C.D, Intersection patterns of curves,
J. London Math. Soc. (2008)

\bibitem{FP4}
J. Fox and J. Pach, 
Computing the independence number of intersection graphs. 
SODA 2011, 1161-1165.



\bibitem{Harry}
   Harry B. Hunt III,  Marathe M. V.,  Radhakrishnan V.,
 Ravi S. S, 
   Rosenkrantz D. J.,  and  Stearns R. E..
 {
   A Unified Approach to Approximation Schemes for NP- and PSPACE-Hard
Problems for Geometric Graphs}.
{ Journal of Algorithms}, 26, 135-149, 1996.

\bibitem{Hoch}
 Hochbaum D.S.,   Maass W.,  {Approximation Schemes
for Covering and Packing Problems in Image Processing and VLSI}.
{ Journal of the Association for Computing Machinery}, 32(1),
 130-136, 1985.

\bibitem{LT}
Lipton, R.J., Tarjan, R.E.,
Applications of a planar separator theorem, SIAM J. Comput., 9(3),615-628,
1980.

\bibitem{LT2} Lipton, R.J.,  Tarjan R.E, A separator theorem for planar graphs, SIAM Journal on Applied Mathematics 36, 1979,  177-189.
.

\bibitem{Mi}
 Miller, G.L.,  Teng, S.,  Thurston W., and  Vavasis, S. A., 
Separators for Sphere-Packings and Nearest Neighborhood graphs, JACM, 44(1),
1-29, 1997

\bibitem{SW}
Smith and N.C. Wormald, Geometric separator theorems and applications. In  39th Annual Symposium on Foundations of Computer Science: FOCS '98, pages 232-243, Palo Alto, CA, 1998.

\bibitem{Sh}
 Shahrokhi F., A New Separation  Theorem with Geometric
Applications, EuroCG2010.










\end{thebibliography}
\end{document}